\documentclass[journal,twocolumn]{IEEEtran}

\usepackage[utf8]{inputenc}
\usepackage{amsmath,amssymb}
\usepackage{graphicx}
\usepackage{float}
\usepackage{hyperref}

\usepackage{booktabs}

\title{Optimizing Neural Architectures for Hindi Speech Separation\\
  and Enhancement in Noisy Environments}
\author{%
  \IEEEauthorblockN{Arnav Ramamoorthy}
  \IEEEauthorblockA{Department of Computer Science\\
    Birla Institute of Technology and Science Pilani, Pilani 333031, India\\
    f20220007@pilani.bits-pilani.ac.in}
}
\date{}  % IEEEtran will handle the date

\begin{document}
\maketitle

\begin{abstract}
This paper addresses the challenges of Hindi speech separation and enhancement using advanced neural network architectures, with a focus on edge devices. We propose a refined approach leveraging the DEMUCS model to overcome limitations of traditional methods, achieving substantial improvements in speech clarity and intelligibility. The model is fine-tuned with U-Net and LSTM layers, trained on a dataset of 400,000 Hindi speech clips augmented with ESC-50 and MS-SNSD for diverse acoustic environments. Evaluation using PESQ and STOI metrics shows superior performance, particularly under extreme noise conditions. To ensure deployment on resource-constrained devices like TWS earbuds, we explore quantization techniques to reduce computational requirements. This research highlights the effectiveness of customized AI algorithms for speech processing in Indian contexts and suggests future directions for optimizing edge-based architectures.
\end{abstract}


% … rest of your sections …

\section{Introduction}
The surge in consumer audio devices like smartphones, tablets, and wireless earbuds has amplified the demand for high-quality audio, especially in noisy environments where traditional methods struggle \cite{wang2018supervised}. Speech enhancement and environmental noise cancellation are crucial for recovering clear speech from recordings marred by ambient noise and reverberation. Traditional speech enhancement methods, such as linear filters and wavelet transforms, offer limited noise reduction and can distort speech \cite{loizou2013speech}. The advent of deep learning, particularly deep neural networks (DNNs) and convolutional neural networks (CNNs), has improved noise reduction while preserving speech fidelity. This research focuses on developing AI-powered speech enhancement algorithms for local Indian languages, specifically Hindi.

Recent advancements in speech separation have seen significant contributions not only from DEMUCS but also from architectures such as Conv-TasNet \cite{Luo2019ConvTasNet}. While Conv-TasNet has demonstrated notable performance in real-time speech separation tasks due to its efficient temporal convolutional network (TCN) design, our study focuses on DEMUCS owing to its superior performance under extreme noise conditions and its compatibility with quantization techniques for edge deployment.

\section{Literature Review}
Traditional speech enhancement methods, like spectral subtraction and Wiener filtering, rely on statistical differences between speech and noise to estimate clean signals but struggle in non-stationary noise environments, leading to a trade-off between noise suppression and speech distortion \cite{loizou2013speech}. Deep learning has advanced the field by capturing complex, non-linear relationships within audio data. Early DNNs mapped noisy speech spectra to clean estimates but had difficulties with temporal dependencies \cite{xu2014experimental}. CNNs improved on this by leveraging the spectro-temporal structure of speech signals, though they were limited in long-range temporal modeling.

RNNs, particularly LSTMs, effectively capture temporal dynamics but often suffer from vanishing gradients. The U-Net architecture, initially developed for image segmentation, has been adapted successfully for speech enhancement with its symmetric encoder-decoder structure and skip connections \cite{kolbaek2016speech}. The DEMUCS model, an adaptation of U-Net with dilated convolutions, further enhances real-time speech processing \cite{demucs2019demucs,demucsICLR2019,demucs2021}.

\textbf{Comparison of Architectures:} Recent advancements in speech separation have seen the emergence of distinct neural architectures that address the problem from varied perspectives. Conv-TasNet \cite{Luo2019ConvTasNet} operates directly in the time domain, while DEMUCS integrates CNN and LSTM components to capture both spectral and temporal features. Hindi speech presents unique challenges due to complex phonetics and pitch variations, which traditional models trained on English data often fail to address. To overcome this, we use a dataset of 400,000 Hindi speech clips augmented with ESC-50 and MS-SNSD to ensure robust training across diverse conditions.

\section{Architecture of the DEMUCS Model}
The DEMUCS (Deep Extractor for Music Sources) model, crafted by Facebook AI Research, is engineered for the task of music source separation and is particularly adept at isolating individual audio components from composite tracks \cite{demucs2019demucs}. At its core, the DEMUCS model incorporates a sophisticated architecture combining convolutional neural networks (CNNs) and recurrent neural networks (RNNs), designed to process both the spectral and temporal dimensions of audio data effectively.

\begin{figure}[h]
    \centering
    \includegraphics[height=0.35\textheight]{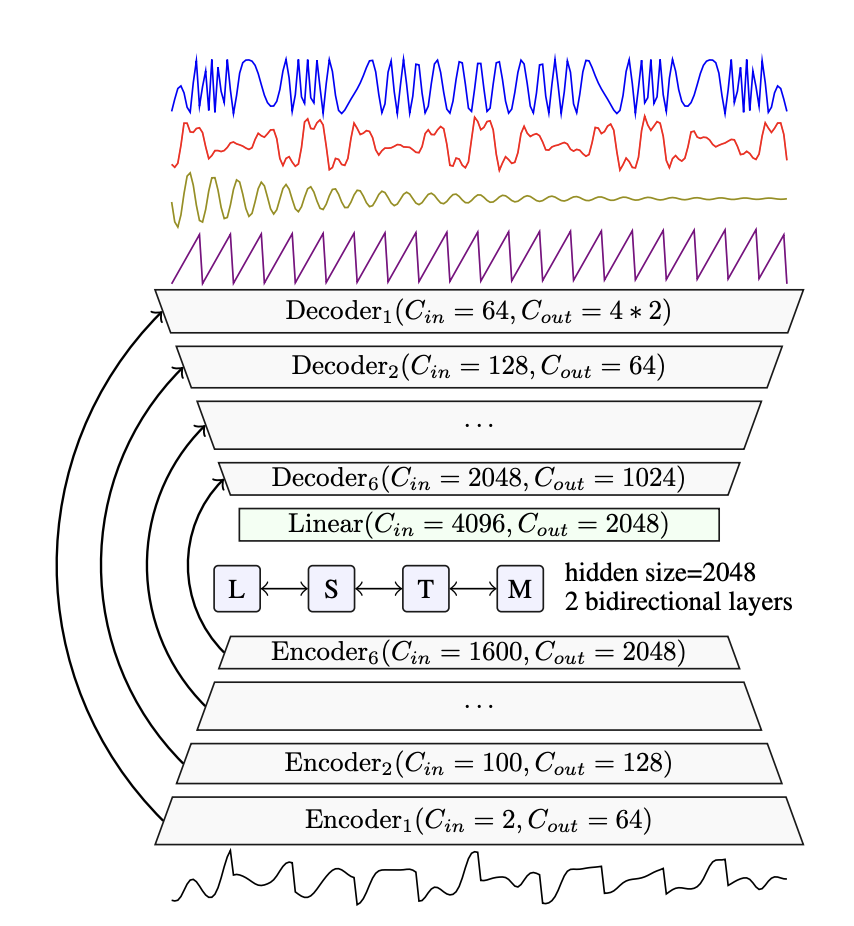} % Adjust the height as needed
    \caption{Diagram of the DEMUCS model architecture}
    \label{fig:demucs_architecture}
\end{figure}

\subsection{Input Transformation}
The model processes the input audio signal, \( x(t) \), through the Short-Time Fourier Transform:
\[
X(f, \tau) = \int_{-\infty}^{\infty} x(t) w(t-\tau) e^{-j2\pi ft} \, dt,
\]
where \( w(t) \) is a window function, \( f \) denotes frequency, and \( \tau \) represents time.

\subsection{Encoder-Decoder Framework}
\begin{itemize}
    \item \textbf{Encoder:} A series of convolutional layers extract and compress spectral features via:
    \[
    E(x) = \sigma\left((x \ast h) + b\right),
    \]
    where \( h \) is the convolutional kernel, \( b \) is the bias, and \( \sigma \) is an activation function (e.g., ReLU) \cite{demucs2019demucs}.
    \item \textbf{Bottleneck and Processing:} The encoded features are refined at the bottleneck to optimize the audio representation for separation.
    \item \textbf{Decoder:} Transposed convolutions reconstruct the separated audio:
    \[
    D(y) = \sigma\left((y \ast h') + b'\right),
    \]
    with \( h' \) as the decoder kernel \cite{demucs2021}.
\end{itemize}

\subsection{Temporal Dynamics Handling}
Integration of Long Short-Term Memory (LSTM) networks enables effective handling of sequential data, preserving essential temporal relationships.

\subsection{Fine-Tuning and Loss Function}
The model is fine-tuned for Hindi speech separation with a learning rate of \(1 \times 10^{-4}\), 10 epochs, and a batch size of 16. A custom loss function, \textbf{ComplexLoss}, is employed:
\[
\mathcal{L} = \alpha \cdot \text{MSE} + \beta \cdot \text{MAE} + \gamma,
\]
with \(\alpha = 0.5\), \(\beta = 0.3\), and \(\gamma = 0.2\). This loss function balances the reduction of large errors with overall consistency and is validated across varying noise conditions.

\section{Dataset}
\subsection{MS-SNSD Hindi Speech Clips}
The primary component of our dataset is derived from the Microsoft Scalable Noisy Speech Dataset (MS-SNSD), comprising 400,000 Hindi speech clips \cite{reddy2019scalable}. These clips are merged into 15-second segments, providing a rich resource for linguistic analysis and speech processing.

\subsection{ESC-50 Environmental Sounds Dataset}
We augment our Hindi speech data with the ESC-50 dataset, a collection of 2,000 environmental sound recordings across 50 classes \cite{piczak2015esc}.

\subsection{Noisy Clip Generation}
Noisy clips are generated by mixing clean Hindi speech from MS-SNSD with noise from ESC-50:
\[
z_k = x_i + \alpha_k n_j, \quad \text{with} \quad \alpha_k = \sqrt{\frac{\sum x_i^2}{\sum n_j^2 \cdot 10^{\text{SNR}_k / 10}}},
\]
where \(x_i\) is a clean speech segment and \(n_j\) is a noise segment. The SNR levels are computed as:
\[
\text{SNR}_k = \text{SNR}_{\text{min}} + \frac{k-1}{L-1} \cdot (\text{SNR}_{\text{max}} - \text{SNR}_{\text{min}}), \quad k = 1,2,\ldots,L.
\]

\section{Results}
Below are three tables comparing the PESQ and STOI metrics for both Conv-TasNet and DEMUCS architectures under overall conditions, 40 dB SNR, and 0 dB SNR. In each table, the header uses a two-lined format indicating the state (Base or Fine-Tuned) and the model name.

\begin{table}[h]
\centering
\caption{PESQ and STOI Median Comparison: Baseline vs. Fine-Tuned (Overall)}
\label{tab:overall}
\begin{tabular}{|l|c|c|c|c|}
\hline
\textbf{Metric} & \textbf{\shortstack{Base\\Conv-TasNet}} & \textbf{\shortstack{Fine-Tuned\\Conv-TasNet}} & \textbf{\shortstack{Base\\DEMUCS}} & \textbf{\shortstack{Fine-Tuned\\DEMUCS}} \\
\hline
PESQ & 1.1 & 1.61 & 1.2 & 1.89 \\
STOI & 0.52 & 0.71 & 0.73 & 0.83 \\
\hline
\end{tabular}
\end{table}

\begin{table}[h]
\centering
\caption{SNR 40 Median Comparison: Baseline vs. Fine-Tuned}
\label{tab:snr40}
\begin{tabular}{|l|c|c|c|c|}
\hline
\textbf{Metric} & \textbf{\shortstack{Base\\Conv-TasNet}} & \textbf{\shortstack{Fine-Tuned\\Conv-TasNet}} & \textbf{\shortstack{Base\\DEMUCS}} & \textbf{\shortstack{Fine-Tuned\\DEMUCS}} \\
\hline
PESQ @ 40 dB & 1.84 & 2.23 & 2.0 & 2.4 \\
STOI @ 40 dB & 0.62 & 0.88 & 0.82 & 0.93 \\
\hline
\end{tabular}
\end{table}

% \begin{table}[h]
% \centering
% \caption{SNR 0 Median Comparison: Baseline vs. Fine-Tuned}
% \label{tab:snr0}
% \begin{tabular}{|l|c|c|c|c|}
% \hline
% \textbf{Metric} & \textbf{\shortstack{Base\\Conv-TasNet}} & \textbf{\shortstack{Fine-Tuned\\Conv-TasNet}} & \textbf{\shortstack{Base\\DEMUCS}} & \textbf{\shortstack{Fine-Tuned\\DEMUCS}} \\
% \hline
% PESQ @ 0 dB & 1.05 & 1.2 & 1.1 & 1.3 \\
% STOI @ 0 dB & 0.43 & 0.64 & 0.61 & 0.73 \\
% \hline
% \end{tabular}
% \end{table}

\begin{table*}[htbp] % Use table* for double column spanning, [htbp] for placement
\centering
\caption{\fontsize{8}{10}\selectfont{{ SNR 0 MEDIAN COMPARISON: BASELINE VS. FINE-TUNED}}} % IEEE caption format
\label{tab:snr0}
\begin{tabular}{|l|c|c|c|c|}
\hline
\textbf{Metric} & \textbf{Base Conv-TasNet} & \textbf{Fine-Tuned Conv-TasNet} & \textbf{Base DEMUCS} & \textbf{Fine-Tuned DEMUCS} \\
\hline
PESQ @ 0 dB & 1.05 & 1.2 & 1.1 & 1.3 \\
STOI @ 0 dB & 0.43 & 0.64 & 0.61 & 0.73 \\
\hline
\end{tabular}
\end{table*}

The analysis clearly demonstrates that fine-tuning significantly improves both speech quality (PESQ) and intelligibility (STOI) across various noise levels, with the DEMUCS architecture showing consistent superiority over Conv-TasNet.

\section{TWS Deployment Experiments}
To evaluate the deployment performance of the DEMUCS-based Hindi speech separation model on resource-constrained True Wireless Stereo (TWS) earbuds, experiments were conducted on a Snapdragon S24 device via Qualcomm AI Hub. The objective was to assess the impact of dynamic INT8 quantization on inference latency and memory footprint.

\subsection{Experimental Setup and Methodology}
The experiment was run across 1200 audio clips from the Hindi speech dataset in batches of 10 clips to simulate real-world conditions. The experimental procedure was as follows:
\begin{enumerate}
    \item \textbf{Baseline Measurement:} The original FP32 DEMUCS model was loaded and evaluated using a custom routine that averaged the inference time (557.87~ms) over multiple runs. Memory usage was determined by tracking resident memory before and after inference.
    \item \textbf{Dynamic INT8 Quantization:} A duplicate model was quantized dynamically using PyTorch's \texttt{torch.quantization.quantize\_dynamic()} method. Selected layers (Linear, Conv1d, and Conv2d) were converted to INT8, reducing the model's memory footprint while computing activation scales at runtime \cite{jacob2018quantization,krishnamoorthi2018quantizing}.
    \item \textbf{Performance Evaluation:} Both models were compared based on inference time and memory usage. Audio quality metrics (PESQ and STOI) were computed to ensure that the quantization did not degrade performance beyond acceptable limits.
\end{enumerate}

\subsection{Results and Analysis}
The quantized model achieved a 1.09$\times$ speedup (514.08~ms compared to 557.87~ms) and a 40.91\% reduction in memory usage (6.58~MB vs.\ 9.25~MB). Although the speedup is modest due to the overhead of dynamically computing activation parameters, the significant memory savings are critical for deployment on low-power, resource-constrained devices.

\section{Conclusion}
This study demonstrates the effectiveness of advanced neural network architectures for Hindi speech separation and enhancement. The experimental evaluations indicate that fine-tuning significantly improves the DEMUCS model, with overall PESQ increasing from 1.2 to 1.89 and STOI from 0.73 to 0.83, as well as notable improvements across different SNR levels. Additionally, dynamic INT8 quantization reduced the model’s memory footprint by 40.91\% and achieved a modest 1.09$\times$ speedup, which is critical for real-time processing on TWS earbuds.

However, the quantization approach introduces runtime overhead due to the dynamic computation of activation scales. Future work should explore static quantization or quantization-aware training (QAT) to further improve inference speed. Additionally, investigating hardware-specific optimizations, such as leveraging Qualcomm’s QNN targets, may overcome current limitations. These enhancements are essential for robust deployment on resource-constrained edge devices and for further advancing Hindi speech processing in challenging acoustic environments.

\bibliographystyle{IEEEtran}
\bibliography{ref}
\end{document}